\documentclass[aps,prb,twocolumn,superscriptaddress,groupedaddress,showpacs]{revtex4-1}
\usepackage{graphicx,xcolor}
\usepackage{subfigure}

\def\K{{\bf K}}

\def\kt{\tilde{\bf k}}

\begin{document}

\title{The Calculation of Two-Particle Quantities in the Typical Medium Dynamical Cluster Approximation}

\author{Y.\ Zhang}
\email{zhangyiphys@gmail.com}
\affiliation{Department of Physics and Astronomy, Louisiana State University,
Baton Rouge, Louisiana 70803, USA}
\author{Y. F.\ Zhang}
\affiliation{Department of Physics and Astronomy, Louisiana State University,
Baton Rouge, Louisiana 70803, USA}
\author{S. X.\ Yang}
\affiliation{Department of Physics and Astronomy, Louisiana State University,
Baton Rouge, Louisiana 70803, USA}
\author{K.-M.\ Tam}
\affiliation{Department of Physics and Astronomy, Louisiana State University,
Baton Rouge, Louisiana 70803, USA},
\author{N. S. Vidhyadhiraja} 
\affiliation{Theoretical Sciences Unit, 
Jawaharlal Nehru Centre for Advanced 
Scientific Research, Bangalore-560064, India}
\author{M.\ Jarrell}
\affiliation{Department of Physics and Astronomy, Louisiana State University,
Baton Rouge, Louisiana 70803, USA}
\affiliation{Center for Computation and Technology, Louisiana State University,
Baton Rouge, Louisiana 70803, USA}

\date{\today}
\begin{abstract}
    The mean-field theory for disordered electron systems without interaction is widely and successfully used to describe equilibrium properties of materials over the whole range of disorder strengths. However, it fails to take into account the effects of quantum coherence and information of localization. Vertex corrections due to multiple back-scatterings may drive the electrical conductivity to zero and make expansions around the mean field in strong disorder problematic.  Here, we present a method for the calculation of two-particle quantities which enable us to characterize the metal-insulator transitions (MIT) in disordered electron systems by using the Typical Medium Dynamical Cluster Approximation (TMDCA). We show how to include vertex corrections and the information about the mobility edge to the typical mean-field theory. We successfully demonstrate the application of the developed method by showing that the conductivity formulated in this way properly characterizes the MIT in disordered systems.

\end{abstract}


\maketitle

\section{Introduction}

Mean field theories, including the Coherent Potential Approximation (CPA)\cite{p_soven_67,taylor_67} and its cluster extensions such as the Dynamical Cluster Approximation (DCA)\cite{m_jarrell_01a}
have been very successful in the description of disordered systems, including the main features of the single particle spectra\cite{s_kirpatrick_70}. They are based upon the mapping of a disordered lattice problem onto an impurity or small periodic cluster embedded in a self-consistently determined effective medium. These theories use the arithmetically averaged spectra to determine the effective medium; however, on average a disordered system without a gap is always a metal.  Thus they always favor the metallic state and cannot capture the physics of electron localization.\cite{t_nakayama_03,d_thouless_74,d_thouless_70,a_mirlin_94,a_lagendijk_09,m_janssen_98,j_brndiar_06, v_dobrosavljevic_03}

However, near the localization transition, the distribution of the density of states becomes highly skewed log-normal distribution~\cite{a_rodriguez_10,
a_mirlin_94,k_byczuk_05}, over many order of magnitude\cite{g_schubert_10}.  Here, the typical value is given by the geometric mean which also vanishes at the transition, giving an order parameter for the transition.  The typical medium theory (TMT) replaces the average spectrum with the typical spectrum to define the effective medium\cite{v_dobrosavljevic_03}. The Typical Medium DCA (TMDCA)\cite{tmdca2} is a cluster extension of the TMT which also accounts for multiple bands and complex disorder potentials\cite{y_zhang_15a,y_zhang_16}.  It has proven quite successful in capturing the transition and its single particle spectra.  However, most experimental measurements 
are described by two particle Green's functions, including transport, most X-ray and neutron scattering, NMR, etc.  While conventional mean field theories, such as the CPA and DCA have been extended to include the description of two-particle quantities including vertex corrections\cite{m_jarrell_01c,th_maier_05a}, the TMT and the TMDCA have not.  Here we introduce such a formalism, and apply it to the calculation of the conductivity of the three-dimensional Anderson model.

The paper is organized as follows: in Sec.~\ref{sec:formalism} we briefly describe the formalism and methods used.  In Sec.~\ref{sec:cond} we apply this formalism to the conductivity.  Then we present the numerical results for the conductivity in Sec.~\ref{sec:results}. Finally, in Sec.~\ref{sec:discuss} we summarize and conclude the paper.

\section{Formalism}
\label{sec:formalism} 
We will take the standard approach of defining the order parameter, $m$, and the conjugate field $h$, which we assume enter into the Hamiltonian as $H=H_0 - hM$, where $M$ is the total magnetization operator with $\left\langle M \right\rangle =Nm$ and $H_0$ is the full Hamiltonian of the system in the absence of the field.  The formalism we will derive will be completely general for any $M$, $h$ and $H_0$ describing a non-interacting disordered system.  For example, $h$ could be a magnetic, or electric, field or represent the application of neutrons or xrays.  However, to be concrete, we will apply the formalism to the calculation of the magnetic susceptibility of the simplest model of disordered electron systems which is the Anderson disorder model, described by the following Hamiltonian
\begin{equation}
H_0=-t\sum_{<ij>\sigma}(c_{i\sigma}^{\dagger}c_{j\sigma}+h.c.)+\sum_{i\sigma}v_{i}n_{i\sigma}\,.
\label{eq:H0}
\end{equation}
Here the first term describes the hopping processes with the hopping constant $t$, between sites $i$ and $j$ (here only the nearest-neighbor hopping is included) with spin $\sigma$, and the second term describes the static scattering processes on the local disorder center. The local potential $v_{i}$ is a site-independent random quantity, with a uniform box disorder distribution, $p(v)={\displaystyle \frac{1}{2W}\Theta(W-|v|)}$.　In our calculation, we use 4$t$ as the energy unit.

\begin{figure}[t]
    \centerline{\includegraphics[scale=0.15]{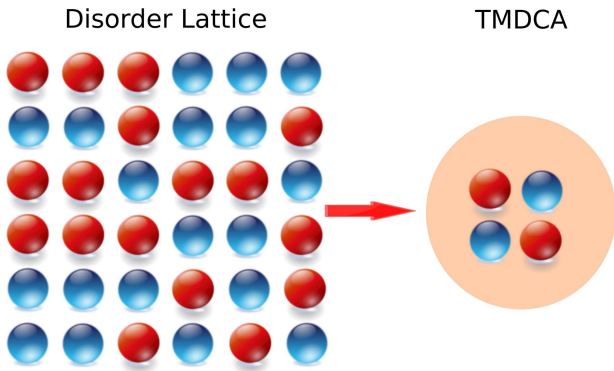}}
\caption{
    Cartoon showing the general idea of the TMDCA.   Instead of solving the original lattice disorder system directly (left), the TMDCA maps the lattice problem to a periodic finite sized cluster embedded in a self-consistently determined typical mean-field (right).  We can calculate the susceptibility of the system by applying the field $h$ to either the lattice problem on the left or the embedded cluster problem on the right. 
}
\label{fig:tmdca}
\end{figure}

To solve this Anderson disorder model, one could work on the lattice  directly. This approach is straightforward but severely limited by the increase in computational cost with the lattice size. Instead, we will employ the typical medium dynamical cluster approximation (TMDCA). Within the TMDCA, the original lattice, of size $N$, is mapped to a finite sized periodic cluster, of size $N_c$, embedded in the self-consistently determined typical mean-field. The mapping is accomplished by averaging the propagators that make up the self energy and irreducible vertex in momentum space by summing over the momenta $\kt$ that surround each cluster momenta $\K$.  See Fig. \ref{fig:tmdca} for a pictorial description of TMDCA. A more detailed discussion of the algorithm has been presented in Ref.~\onlinecite{y_zhang_16}.  The essential difference of the TMDCA as compared to the conventional DCA\cite{m_jarrell_01a,m_jarrell_01c,th_maier_05a} is the typical disorder averaging used when solving the referenced cluster problem. 

One may calculate the order parameter $m$ either in the original lattice problem or in the TMDCA embedded cluster problem.  Here, we will start with the former.  It is important when calculating the susceptibility to apply the field $h$ to every site in the system to assess the linear response.  To this end, we will define the order parameter using the Green's function in lattice wavenumbers
\begin{equation}
m = \frac{1}{N} \sum_{k,\omega,\sigma} \sigma G_{l,\sigma}(k,\omega)\,.
\label{eq:OPlattice}
\end{equation}
Here, $N$ is the total number of lattice sites and implicit in the sum over $\omega$ is a convergence factor which causes the contour to be closed in the lower half plane and the lattice time ordered Greens function is defined as
\begin{equation}
G_{l,\sigma}(k,\omega) = \frac{1}{\omega - h\sigma - \epsilon_k - \Sigma(M(k),\omega)}
\end{equation}
where $M(k)=K$ maps an arbitrary wave number $k$ to the closest DCA cluster $K$ and $\Sigma(M(k),\omega)$ is the time-ordered self energy calculated on the cluster.
To calculate the susceptibility, $\chi = dm/dh|_{h=0}$ it is important to know the derivative of  $\Sigma$ with $h$ since, as we will see below, it is the origin of the vertex corrections.  This representation works well for states below the energy (the mobility edge) where all of the states are extended, since they may be described as states with a dispersion $ \epsilon_k$ renormalized by $\Sigma$.  However, for localized states, above the mobility edge,
the electrons are localized to the cluster and may not be described as extended states with a renormalized dispersion.  So, it makes more sense to choose the other representation, of the periodic cluster embedded in an effective medium
\begin{equation}
m = \frac{1}{N_c}\sum_{K,\omega,\sigma} \sigma G_{c,\sigma}(K,\omega)
\label{eq:OPcluster}
\end{equation}
where $G_{c,\sigma}$ is the TMDCA average cluster Green's function
\begin{equation}
G_{c,\sigma}(K,\omega) = \frac{1}{\omega - h\sigma - \bar \epsilon_K - \Delta_\sigma(K,\omega) - \Sigma_\sigma(K,\omega)}\,,
\end{equation}
$\bar \epsilon_K$ is the DCA coarse grained dispersion $\bar \epsilon_K = \frac{N_c}{N} \sum_{\tilde{k}} \epsilon_{K+\tilde{k}}$,
$\Delta_\sigma$ is the hybridization between the cluster and the effective medium for spin $\sigma$, and the sum on $K$ is over the wavenumbers of the periodic DCA cluster.  
Again, for the calculation of the susceptibility, it is important that we also know the derivatives $d \Sigma_\sigma/dh|_{h=0}$ and now also $d \Delta_\sigma/dh|_{h=0}$.  However, for the localized states above the mobility edge, $\Delta_\sigma=0$ so the states are localized to the cluster, and so $d \Delta_\sigma/dh|_{h=0} = 0$.  
The lack of knowledge of this derivative also prevents us from using this representation for the extended states where $\Delta_\sigma$ is finite.

One may combine these two perspectives by defining a physical Green's function
\begin{equation}
G_{p,\sigma}(k,\omega) = \left\{ \begin{array}{ll}
         G_{l,\sigma}(k,\omega) & \mbox{if $ |\omega| < \omega_e$};\\
         G_{c,\sigma}(M(k),\omega) & \mbox{if $|\omega| > \omega_e$}.\end{array} \right.
         \label{eq:Gp}
\end{equation}
where $\omega_e$ is the mobility edge energy.  Then,
\begin{equation}
m= \frac{1}{N}\sum_{k,\omega,\sigma} \sigma G_{p,\sigma}(k,\omega)
\end{equation}
and the spin susceptibility is 
\begin{equation}
\chi_{s} = \sum_{k,\omega,\sigma} \sigma \left. \frac{d G_{p,\sigma}(k,\omega)}{dh}\right|_{h=0}
\end{equation}
Note that the derivative with $h$ breaks an internal line with labels which we will later identify as $k',\omega',\sigma'$ in Eq.~(\ref{eq:Gtochi}). These labels are summed over, since they are internal labels in the Green function.   So, we can use an implied matrix representation in these indices, 
\begin{eqnarray}
\frac{d G_{p,\sigma}}{dh} &=& \frac{d}{dh}\left( {G_{p,\sigma}^{0}}^{-1} -\Sigma_\sigma \right)^{-1}\nonumber\\
&=& -G_{p,\sigma}^2 \left( \frac{d{G_{p,\sigma}^{0}}^{-1} }{dh} - \frac{d\Sigma_\sigma}{dh} \right)\nonumber\\
&=& G_{p,\sigma}^2 \left( \sigma + \frac{d\Sigma_\sigma}{dh} \right)
\end{eqnarray}
where $G_{p,\sigma}^{0}$ is the bare physical Green's functions (i.e., with $\Sigma = 0$).  The self energy $\Sigma_\sigma $ depends on $h$ only through the single particle Greens function, so
\begin{equation}
\frac{d G_{p,\sigma}}{dh}  
=
G_{p,\sigma}^2 \left( \sigma + \frac{d\Sigma_\sigma}{dG_{p,\sigma''}}
\sigma''^2
\frac{dG_{p,\sigma''}}{dh} \right)
\label{eq:BSE0}
\end{equation}
where $\sigma''^2 =1$.  The irreducible vertex enters through the term 
\begin{equation}
\frac{d\Sigma_\sigma(k,\omega)}{dG_{p,\sigma'}(k',\omega')} = 
\Gamma_{\sigma,\sigma'}(k,\omega; k',\omega') 
\end{equation}
which we will also represent as a matrix. 

We may also write $\chi$ in terms of the density-density correlation functions
\begin{equation}
\chi_{\sigma,\sigma'}(t,i,j) = -i \left\langle T n_\sigma (i, t) n_{\sigma'}(j,0) \right\rangle
\end{equation}
so that
\begin{equation}
\chi_s= \sum_{k,w,\sigma;k',\omega',\sigma'} \sigma \chi_{\sigma,\sigma'}(k,\omega;k',\omega')\sigma'
\end{equation}
So, we may identify
\begin{equation}
\sigma \left. \frac{d G_{p,\sigma}(k,\omega)}{dh}\right|_{h=0} = \sum_{k',\omega',\sigma'}\sigma \chi_{\sigma,\sigma'}(k,\omega;k',\omega') \sigma'
\label{eq:Gtochi}
\end{equation}
for which we will also use a matrix representation in the frequency and wavenumber labels.  So, Eq.~(\ref{eq:BSE0}) becomes
\begin{equation}
\sigma \chi_{\sigma,\sigma'}\sigma'=
\sigma \chi^0\sigma +
\sigma \chi^0 \sigma \sigma \Gamma_{\sigma,\sigma''}\sigma''
\sigma'' \chi_{\sigma'',\sigma'} \sigma'
\end{equation}
which is the Bethe-Salpeter equation, which differs from the conventional definition only through the replacement of $G_l$ by $G_p$.

Note that we can also calculate the bulk ($q=0$) charge susceptibility used in the next section, as
\begin{equation}
\chi_c= \sum_{k,w,\sigma;k',\omega',\sigma'} \chi_{\sigma,\sigma'}(k,\omega;k',\omega')
\end{equation}

Similar steps may be used to derive a Beth-Salpeter equation for the charge, pair and general spin (i.e., including, e.g., the antiferromagnetic) susceptibility in the usual way where we start with a definition of the order parameter, Eqs.~(\ref{eq:OPlattice}) and (\ref{eq:OPcluster}) in terms of the single particle Green's function \cite{m_jarrell_01a,m_jarrell_01c,th_maier_05a}.  In these cases one generally introduces a spinor notation, e.g., the Nambu spinors to calculate the pairing susceptibility.  We then take the derivative with respect to to the conjugate applied field and take the limit as the field goes to zero to obtain the corresponding Bethe-Salpeter equation.

Here, since the two-particle quantities corresponding to the physically
measurable quantities are defined through the algebraic average, the self-energy
defined in the derivations above is calculated by the the algebraic average
after the TMDCA iterations attain convergence, where we use the typical 
hybridization function as the effective medium.

\section{Calculation of the Conductivity}
\label{sec:cond}
\begin{figure}[t]
    \centerline{\includegraphics[scale=0.85]{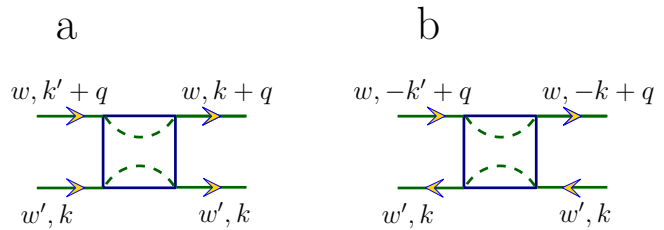}}
\caption{
    Representation of the two-particle quantities in the particle-hole (a)
    and the particle-particle channel (b). Note that for disorder systems,
    only elastic scatterings exist, and therefore only two frequencies are needed 
    (in the figure, following the dash lines, the energy is conserved). 
    For the retarded-advanced component, we assign the above line 
    $\omega$ as the retarded Fermionic line and the lower line $\omega'$ as the 
    advanced Fermionic line. In addition, we focus mainly on the DC conductivity 
    at zero temperature, so we only need zero frequencies, 
    namely $\omega=0$ and $\omega'=0$, for all the two-particle quantities involved. 
    This would greatly simplify our two-particle calculation.
}
\label{fig:vertex} 
\end{figure}

In this section, as an example, we will apply this formalism to study the behavior of DC conductivity at zero temperature, and use it to identify the critical disorder strength where Anderson localization happens. 

To fix the representation of two-particle quantities used for the DC conductivity, Fig.~\ref{fig:vertex} shows the convention we use throughout our calculation\cite{rammer2004quantum,altland2010condensed}. Note that we need three indices for the momentum, while we only need two indices for the energy, which is a consequence of the property of the disorder system where all the scatterings are elastic and following each Fermionic line the energy should be conserved. For the retarded-advanced component, we assign the above line $\omega$ as the retarded Fermionic line and the lower line $\omega'$ as the advanced Fermionic line.

The DC conductivity can be calculated according to the lattice dynamical susceptibility. At zero temperature, it can be expressed as
\begin{equation}
    \sigma_{dc} = \frac{1}{\pi N^2} \sum_{k,k'} v_k [\chi^{ra}+\chi^{rr}](q=0;\omega=0,\omega'=0)_{k,k'} v_{k'}
\label{eq:cond} 
\end{equation}
with the velocity form-factor (to be concrete here we choose the $k_x$ component) defined as
\begin{equation}
    v_k = 2 t \sin(k_x).
\end{equation}
Note that we only need zero frequencies, namely $\omega=0$ and $\omega'=0$, 
for the dynamic charge susceptibility. This special property also applies to other two-particle quantities involved and therefore can be used to greatly simplify our two-particle calculation. To simplify the expression 
in the following, we hide the explicit dependence of 
this trivial frequency dependence. 

The lattice susceptibility is calculated from the lattice full vertex.
In order to do so, we approximate the irreducible lattice vertex function
by the one calculated in the cluster, which is the standard approximation in 
DCA when calculating the two particle quantities\cite{m_jarrell_01c,th_maier_05a}.
So to calculate the conductivity, we need to measure cluster two-particle Green functions after the TMDCA iterations attain 
convergence. Specifically, we need to measure both the r-a and r-r components of the two-particle Green functions as
\begin{equation}
    \chi_c^{ra}(Q)_{K,K'} = \langle g_c^r(K+Q,K'+Q) g_c^a(K',K) \rangle_{dis}
\label{eq:chic_ra_ave}
\end{equation}
and
\begin{equation}
    \chi_c^{rr}(Q)_{K,K'} = \langle g_c^r(K+Q,K'+Q) g_c^r(K',K) \rangle_{dis}.
\label{eq:chic_rr_ave}
\end{equation}
In the above, $\langle...\rangle_{dis}$ represents the averaging over the disorder configurations, and $g_{c}^{r}$ ($g_{c}^{a}$) is the single-particle 
retarded (advanced) cluster Green function for each disorder configuration.
In the following, we use the r-a component as the example and 
the r-r component should follow similarly.
Since the two-particle Green functions are calculated exactly on the cluster,
they include all the diagrams contributing to the full vertex within the cluster.
The cluster full vertex functions are extracted as
\begin{equation}
    F_c^{ra}(Q)_{K,K'} = \frac{\chi^{ra}_c(Q)_{K,K'} - G^r(K+Q)G^a(K)\delta_{K,K'}}
        {[G^r(K+Q)G^a(K)]^2}
\end{equation}

Next, to calculate the lattice full vertex, we could use either Bethe-Salpeter (B.S.) equation, or the following form, which is also exact and motivated by the dual-fermion formalism\cite{a_rubtsov_08,h_terletska_13}, 
\begin{eqnarray}
 F^{ra}(q=0)_{kk^{\prime}} = \left[\frac{F_c^{ra}}{1+F_c^{ra}\chi^{0ra}_{d}}\right](q=0)_{KK^{\prime}}.
\end{eqnarray}
which is more numerically stable. Here,
\begin{equation}
\chi^{0ra}_{d} = \chi_c^{0ra} - \overline{\chi_p^{0ra}}, 
\end{equation}
where $\chi_c^{0ra}$ and $\chi_p^{0ra}$ are the bubble for the cluster Green function and the previously defined physical Green function and the overbar denotes an average within the coarse-grained cells.

Given the lattice full vertex, in the end, we can calculate the lattice susceptibility as ($G^{r/a}_p$ in the following are the physical single-particle Green function discussed in the previous section)
\begin{eqnarray}
    &&  \chi^{ra}(q=0)_{k,k'}  \\ 
    &=& G^r_p(k)G^a_p(k)\delta_{k,k'} \nonumber\\ 
    &+& G^r_p(k)G^a_p(k)F^{ra}(q=0)_{k,k'}G^r_p(k')G^a_p(k') \nonumber\\
    &=& \chi_p^{0ra}+\chi_p^{0ra} F^{ra}(q=0)\chi_p^{0ra} \nonumber
\end{eqnarray}
which can be readily inserted into Eq.~(\ref{eq:cond}) to get the DC conductivity we need.
When calculating the DC conductivity using above formula, since the lattice full vertex $F^{ra}(q=0)_{k,k'}$ here depends on $k$ and $k'$ only through the cluster momentum $M(k)$ and $M(k')$, an average over the coarse-grained cells can be carried out before the summation over $k$ and $k'$. Then the DC conductivity becomes:
\begin{equation}
    \sigma_{dc} = \frac{1}{\pi N_c^2} \sum_{K,K'} \overline{v [\chi^{ra}+\chi^{rr}]v}(q=0;\omega=0,\omega'=0)_{K,K'} 
\end{equation}
where,
\begin{equation}
\overline{v \chi^{ra} v}=\overline{v \chi_p^{0ra} v}+\overline{v \chi_p^{0ra}} F^{ra}(q=0) \overline{v \chi_p^{0ra}}
\end{equation}
and similarly for r-r component.

\begin{figure}[t]
    \centerline{\includegraphics[scale=0.65]{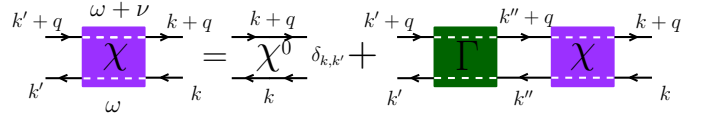}}
\caption{
Bethe-Salpeter equation relating the two-particle Green function $\chi$ and the irreducible vertex $\Gamma$. While $k$, $k'$ and $q$ represent momentum indices, $\omega$ and $\nu$ represent frequency indices (for fermionic and bosonic frequencies respectively). Note that for disorder systems considered here, the scatterings are elastic and thus the energy is conserved following any fermionic Green function line. Therefore, we only need two frequency indices to represent the frequency degree of freedom of the system.
}
\label{fig:bs}
\end{figure}

\section{Results}
\label{sec:results}

\begin{figure}[t]
    \centerline{\includegraphics[scale=0.35]{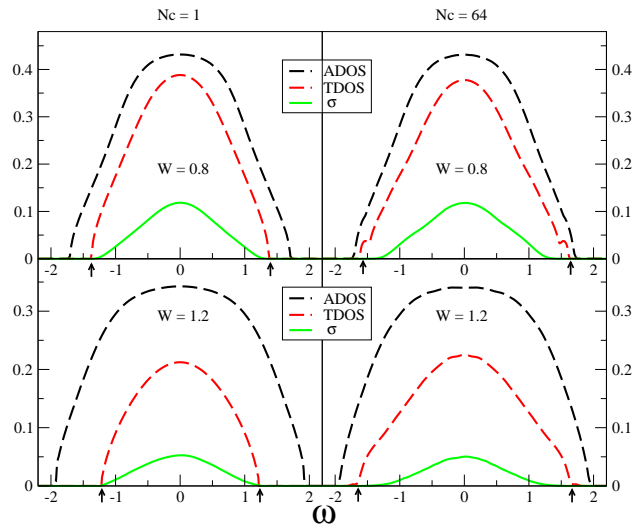}}
\caption{
    The evolution of the ADOS, TDOS and dc conductivity at various disorder strengths $W$ for the single-site TMT and the TMDCA with cluster size $N_c=64$. Here, for dc conductivity, $\omega$ corresponds to the chemical potential used in the calculation. At low disorder,
where all the states are metallic, the shape of TDOS is the same as that of the ADOS.  As $W$ increases, in the case of single-site TMT, the TDOS gets suppressed and the mobility edge moves towards $\omega=0$ monotonically. In the TMDCA,
the TDOS is also suppressed, but the mobility edge first moves to higher energy, and only with a further increase of $W > 1.8$, it starts moving towards the band center,
 indicating that the TMDCA can successfully capture the re-entrance behavior\cite{tmdca2}. Arrows indicate the position of the mobility edge, which separates the extended electronic states from the localized ones and the colored region indicates the TDOS.}
\label{fig:energy}
\end{figure}

First (Fig.~\ref{fig:energy}) we study the dc conductivity and the density of states, including the algebraically averaged DOS (ADOS) and the typical DOS (TDOS), as a function of the chemical potential for N${}_c$ = 1 and 64 at various disorder strengths.   To calculate the DC conductivity, we use the TMDCA to acquire a converged typical hybridization function, which already includes the information of localization, and then we will use this typical hybridization function in and additional TMDCA iteration to calculate the algebraically averaged cluster single and two particle Green functions, e.g., Eqs.~(\ref{eq:chic_ra_ave}) and (\ref{eq:chic_rr_ave}).  
The ADOS is obtained from the conventional DCA scheme, where the ADOS is used in the self-consistency.  As can be seen from Fig.~\ref{fig:energy} for both TMT (N${}_c$ = 1) and TMDCA (N${}_c$ = 64), the ADOS remains finite while the TDOS gets suppressed as the disorder strength $W$ is increased.  From Fig.~\ref{fig:energy}, we can also see that the conductivity vanishes in the region where the TDOS is zero.  This should be the case since when the typical density of state is zero, which means all states are localized on the cluster, the hybridization function also becomes zero and all clusters are isolated.  Just below the mobility edge, where the TDOS is finite, the conductivity is very small, but as we will see next, it remains finite.

The conductivity plotted in Fig.~\ref{fig:energy} is a smooth function of energy, but for very weak disorder strength, e.g. $W=0.25$ (not shown), the conductivity has more structure near the band center, due to the Van Hove singularities in the DOS.  They are not completely destroyed by the relatively weak disorder as shown in Fig.~5.8 of Ref.~\onlinecite{P_Sheng_2006}, and our calculation shows similar behavior for weak disorder strengths. Since we focus on the intermediate and strong disorder cases, where the Van Hove singularity is destroyed by the disorder, we do not reproduce these results here.

\begin{figure}[t]
    \centerline{\includegraphics[scale=0.35]{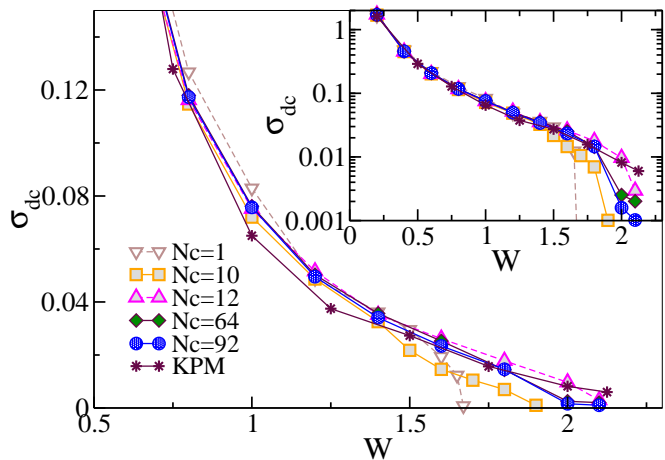}}
\caption{
    DC conductivity of the 3D Anderson model at $T=0$ and $\mu=0$(band center) vs disorder $W$ for different
    cluster size $N_{c}=1,10,12,64,92$. The dc conductivity vanishes at $W_{c}$ where all states become localized.
    For $Nc=1$(TMT), the critical disorder  strength  $W_c^{N_c=1} \approx 1.65$.
As the cluster size increases, $W_c$ systematically increases with
$W_c^{N_c\gg12}\approx 2.10\pm 0.10$, showing a quick convergence with the cluster size. At large enough $N_{c}$,
our results converge to those obtained by KPM.
}
\label{fig:disorder}
\end{figure}

Next, we consider the evolution of $W_c$ with N${}_c$.  Fig.~\ref{fig:disorder} shows the DC conductivity at zero energy 
as a function of $W$ for several N${}_c$. $W_c$ is defined by the vanishing of the
DC conductivity. Our results show that as cluster size N${}_c$ increases, for N${}_c\ge12$ the
$W_c$ systematically increases until it converges
to $W_c\approx 2.1$  which is in
good agreement with the values reported in the literature.  This cluster is the first one with a complete nearest neighbors shell. From this cluster onward, $W_c$ converges to $\approx$ 2.1.
We compare our TMDCA results with the Kernel polynomial method(KPM)\cite{a_weisse_06,Weisse2004,Gracia-Covaci-Rappoport-2015,Leconte-Ferreira-Jung-2016}.  For most values of the disorder strength the agreement with the KPM is excellent. The results get noisy near the transition (Fig.~\ref{fig:disorder}), but the deviation from the KPM calculations is in the correct direction and the conductivity vanishes near the critical disorder strength.

In contrast to the other cluster methods for conductivity, such as the transfer matrix method and KPM, a clear advantage of the present method is that we can incorporate realistic electronic structure as in, e.g., the dynamical mean-field theory~\cite{RevModPhys.68.13} and other DCA~\cite{PhysRevB.73.085106}, and TMDCA\cite{y_zhang_16} calculations.

\section{Discussion and Conclusion}
\label{sec:discuss}
Numerous two-particle quantities of interest can be calculated with this method.  These include quantities such as the RKKY, double or kinetic exchanges in magnetic systems, where were all originally formulated for extended electronic states, 
or the full set of Onsager coefficients and related transport quantities, including the thermal transport.  Also, the methods presented here may easily be extended to the calculation of dynamic spin and charge quantities, such as the optical conductivity, or dynamic spin susceptibility.  In each case, the method is the same as that generally used in DCA, with the replacement of the lattice single-particle propagator with $G_p$.
In addition, similar methods may be used to study realistic model systems with multiple bands and complex disorder potentials\cite{y_zhang_16}.  

This formalism works for non-interacting disordered systems.  The introduction of interactions will introduce inelastic scattering which, at least for weak interactions, destroys the mobility edge until the edge and the chemical potential coincide, at which point the edge again becomes sharp\cite{c_ekuma_15c}.  Without a sharp edge for all values of the chemical potential, one cannot partition the Green function as done in Eq.~(\ref{eq:Gp}).  So, this formalism will fail.  Note that it is possible to formulate a method for interacting systems which relies only upon the response of the embedded cluster problem to an applied field.  However, this method would require that we know how to calculate $d \Delta_\sigma/dh|_{h=0}$. 

For the calculation of the conductivity, the result changes rather abruptly from a finite value to a numerical zero as the mobility edge is crossed.  This is by construction, since the scattering is elastic, the results at different frequencies are independent.  Due to the form of $G_l$, the calculation abruptly changes from the conductivity of the lattice to the conductivity of an isolated periodic cluster, which has zero conductivity, as the mobility edge is crossed.  

It is useful to compare the usefulness of the conductivity and the typical density of state (TDOS) as order parameters for the localization transition. As illustrated in Fig.~\ref{fig:energy}, especially for larger clusters ($N_c=64$ shown on the right), the mobility edge given by the TDOS is sharp and well defined; whereas, the conductivity becomes very small and noisy well below the edge, making it more difficult to identify the mobility edge with the conductivity.  Thus, for numerical calculations such as this, the TDOS is a superior order parameter for the identification of the mobility edge.

Finally, we note that, as illustrated in Fig.~\ref{fig:energy}, the conductivity calculated for small clusters ($N_c=1$) barely differs from that from large clusters ($N_c=64$) until the disorder strength is very close to the transition value.  One difference between these two calculations is that for $N_c=1$ there are no vertex corrections\cite{PhysRevLett.64.1990}, which are present for larger clusters.  Apparently, vertex corrections are not significant in the conductivity until very close to the transition.  

\begin{acknowledgments}
We thank Hanna Terletska for useful discussions and comments. This work is supported by NSF EPSCoR Cooperative Agreement No. EPS-1003897 (YZ, SY, YFZ, KMT, and MJ). This work used the Extreme Science and Engineering Discovery Environment (XSEDE), which is supported by National Science Foundation grant number ACI-1053575. This work also used the high performance computational resources provided by the Louisiana Optical Network Initiative (http://www.loni.org), and HPC@LSU computing.
\end{acknowledgments}

\bibliography{cond_dis}

\end{document}